\documentclass[prl,twocolumn,superscriptaddress,10pt,noshowpacs]{revtex4}
\usepackage[english]{babel}
\usepackage[T1]{fontenc}
\usepackage[utf8]{inputenc}
\usepackage{graphicx,epstopdf}
\usepackage{amsmath}

\usepackage[pdftex,plainpages=false,colorlinks=true,citecolor=blue,linkcolor=blue,urlcolor=blue,filecolor=green,bookmarksopen=true]{hyperref}
\usepackage{amsfonts}
\usepackage{bbm}
\usepackage{amssymb}
\usepackage{xcolor}
\usepackage{latexsym}
\usepackage{times,txfonts}

\newcommand{\nab}{\vec{\nabla}}

\begin{document}
\title{Lorentz-violating QED inspired superconductivity}

\author{A. F. Morais}
\email{apiano.morais@urca.br}
\affiliation{Laborat\'orio de F\'isica Computacional Aplicada, Universidade Regional do Cariri, 63041-235, Juazeiro do Norte, Cear\'{a}, Brazil}

\author{M. C. Ara\'{u}jo}
\email{michelangelo@fisica.ufc.br}
\affiliation{Centro de Ci\^{e}ncias e Tecnologia, Universidade Federal do Cariri, 63048-080, Juazeiro do Norte, Cear\'{a}, Brazil}

\author{T. T. Saraiva}
\email{tteixeirasaraiva@hse.ru}
\affiliation{HSE University, Moscow 101000, Russia}
\affiliation{Centro de Ci\^{e}ncias e Tecnologia, Universidade Federal do Cariri, 63048-080, Juazeiro do Norte, Cear\'{a}, Brazil}

\author{J. Furtado}
\email{job.furtado@ufca.edu.br}
\affiliation{Centro de Ci\^{e}ncias e Tecnologia, Universidade Federal do Cariri, 63048-080, Juazeiro do Norte, Cear\'{a}, Brazil}

\date{\today}

\begin{abstract}
We studied a Lorentz-violating inspired Ginzburg-Landau model for superconductivity where we considered a CPT-odd contribution given by $(k_{AF})^{\mu}$, also known as the Carroll-Field-Jackiw term. In the static limit of the equations, we could find a pair of modified Ginzburg-Landau equations. Furthermore, these equations were reduced to the London equation for the magnetic field when assumed that the characteristic length of the order parameter is much smaller than the characteristic length of the magnetic field, i.e. the London penetration length. Our numerical solutions showed a simple Meissner state when this new term is small compared to $\lambda_L$ and a phase transition into phases with strong in-plane currents and anomalous vortices for large contributions. This model becomes useful in exemplifying the changes in the phenomenology of superconductors when the setup of the system shows an important breakdown of Lorentz invariance. Based on these results, we discuss how such models might be the hallmark of unusual superconducting states where there is a direction where the system shows stratification, as in anapole superconductors UTe$_2$.
\end{abstract}

\maketitle

The Standard Model of Particle Physics (SM) is a well-established theory that appropriately describes three fundamental interactions in nature: electromagnetic, weak, and strong. However, despite its experimental success, the SM still leaves some questions unanswered, such as neutrino oscillations \cite{barger2012physics}, the hierarchy problem \cite{Arkani-Hamed:1998jmv}, and the lack of a quantum description of gravity. To address these issues, there has been significant interest in exploring possible extensions of the SM. Within this new context, the Standard Model Extension (SME) emerges as an alternative approach to investigate measurable signals of Lorentz and CPT symmetry violation at low energies \cite{Colladay:1996iz,Colladay:1998fq}. The fundamental concept behind the SME is to extend the Lagrangian density of the SM to an effective field theory that includes all possible terms violating Lorentz symmetry while still preserving the gauge symmetries present in the SM. Since any parameter violating CPT symmetry also violates Lorentz symmetry, the SME provides a comprehensive framework for studying systems with the potential for CPT symmetry breaking \cite{Kostelecky:1988zi,Colladay:1996iz,Colladay:1998fq}.

As a comprehensive framework for treating CPT violation at the level of effective field theory, the SME has led to several experimental investigations \cite{Link:2002fg,BaBar:2007run,KLOE-2:2013ozx,KLOE:2010yad,D0:2015ycz,LHCb:2016vdl}. Moreover, it has
been extensively studied in numerous contexts such as neutrino oscillation \cite{MINOS:2008fnv,MINOS:2010kat,Katori:2012pe,MiniBooNE:2011pix}, radiative corrections \cite{Jackiw:1999yp,Kostelecky:2001jc, Ferrari:2021eam, Furtado:2014cja}, supersymmetric models \cite{Lehum:2018tpi,Ferrari:2017rwk,Belich:2015qxa,Colladay:2010tx}, and noncommutative quantum field theories \cite{Carroll:2001ws,Hayakawa:1999yt}.

Models that violate Lorentz and/or CPT symmetry have also gained considerable traction in condensed matter physics, with applications in Weyl semimetals \cite{Assuncao:2015lfa, Gos1, Gos2, Gos3}, graphene with anisotropic scaling \cite{Katsnelson:2012cz}, analogue models of dark matter and black holes \cite{Baym:2020uos, Pereira:2009vb}, and superconductivity \cite{Bazeia:2016pra}, to name a few examples. Recently, possible anisotropic effects in superconductors have been investigated through models inspired by the scalar sector of the SME \cite{Furtado:2020iom,Araujo:2024qha}. In these studies, the authors considered a modified Ginzburg-Landau (GL) theory and examined how the symmetry-breaking parameters affect the characteristic lengths of a material, as well as the critical magnetic fields under which the superconducting state is lost. It's worth noting that superconductivity in Weyl semimetals \cite{Zhou:2015qka, Bednik:2015tha, Wei:2014vsa} and carbon-based nanostructures, such as graphene and  fullerene \cite{Roy:2013aya, Cohnitz:2017vsr}, is a topic of frequent interest due to potential technological applications. However, from a more theoretical perspective, the subject can be extended to astrophysical \cite{Madsen:1999ci, Bonanno:2011ch}, cosmological \cite{Ebert:2007ey, Gao:2012aw}, and high-energy physics  \cite{Herzog:2009xv, Ghoroku:2019trx} domains as a whole.

In this article, we propose a modified GL model inspired by a scalar electrodynamics within the SME framework. Starting from a Lagrangian that minimally couples a complex scalar field with the gauge field and includes a CPT-odd modification in the pure gauge sector, we derive a Lorentz-violating Ginzburg-Landau free energy density. This allows us to investigate anisotropic effects on the properties of superconducting materials, such as characteristic lengths and critical magnetic fields~\cite{Tinkham,deGennesBook}. We also analyze scenarios with anomalous behaviors of current and magnetic field inside a superconductor described by a modified Lorentz-violating London equation and how this could be the signature for unconventional superconductors~\cite{Fu2015,Kozii2015,Schumann2020,Kanasugi2022}.

Our starting point in investigating possible Lorentz symmetry violation effects in superconducting materials is the Lagrangian density 
\begin{eqnarray}\label{lagrangian1}
\nonumber\mathcal{L}&=&(D_{\mu}\phi)^*D^{\mu}\phi-\mu^2\phi^*\phi-\lambda(\phi^*\phi)^2-\frac{1}{4}F^{\mu\nu}F_{\mu\nu}\\
&+&\frac{1}{2}\left(k_{AF}\right)^{\kappa}\epsilon_{\kappa\lambda\mu\nu}A^{\lambda}F^{\mu\nu}.
\end{eqnarray} Here, $D_{\mu}=\partial_{\mu}-ieA_{\mu}$ represents the usual covariant derivative, allowing for coupling between the complex scalar sector and the gauge field, and $F_{\mu\nu}=\partial_{\mu}A_{\nu}-\partial_{\nu}A_{\mu}$ corresponds to the usual Maxwell tensor of electromagnetism. Lorentz violation effects are governed by the vector coefficient $\left(k_{AF}\right)^{\kappa}$, which breaks the equivalence between observer and particle transformations, also known as the Carroll-Field-Jackiw (CFJ) term. Initially, this vector is assumed to be constant to ensure translational invariance and, consequently, the conservation of momentum and energy. Note that $(k_{AF})^{\kappa}$ has dimension of mass and is CPT-odd, as we can see in Table \eqref{tablediscretsimetries}.
\begin{table}[t]
\begin{tabular}{c|c|c|c|c}
\hline
 & \,\,C\,\, & \,\,P\,\, & \,\,T\,\, & \,\,CPT\,\, \\ 
\hline
\,\,\,$(k_{AF})_{j}$\,\,\, & + & + & - & - \\ \hline
$(k_{AF})_{0}$ & + & - & + & - \\ 
\hline
\end{tabular}\caption{Analysis of the discrete symmetries of charge conjugation $C$, parity $P$, and time reversal $T$ on the vector coefficient $(k_{AF})^{\kappa}$.}
\label{tablediscretsimetries}
\end{table}
In the static condition, only the temporal component of the Lorentz-violating coefficient contributes non-null values to the system analysis. Consequently, the Lagrangian density in Eq. \eqref{lagrangian1} can be written as 
\begin{eqnarray}\label{lagrangian2}
  \mathcal{L}&=& -\mu^2|\phi|^2-\lambda|\phi|^4 - |(\nabla-ie\mathbf{A})\phi|^2-\frac{1}{2}\left(\nab \times \mathbf{A}\right)^2\nonumber\\
  &+& k\, \mathbf{A}\cdot(\nab \times \mathbf{A}) 
\end{eqnarray}
with $k$ being the zero-component $(k_{AF})^{0}$ from now on. In our notation, $e$ is the elementary charge.

Inspired by this Lagrangian, one can construct the analogous Ginzburg-Landau free energy expansion for the case where one has a violation of the Lorentz symmetry.
The Ginzburg-Landau free energy density difference between the superconducting and normal states (in cgs units) becomes:
\begin{align}\label{Eq.GLfree}
\Delta\mathfrak{f}&=a\tau|\Psi|^2+\frac{b}{2}|\Psi|^4+\mathcal{K}\left|\left(\nab-i\frac{2e}{\hbar c}\mathbf{A}\right)\Psi\right|^2+\frac{1}{8\pi}\left(\nab \times \mathbf{A}\right)^2\nonumber\\
&\quad-\frac{k}{4\pi}\, \mathbf{A}\cdot(\nab \times \mathbf{A})
\end{align}
with new Ginzburg-Landau coefficients, i.e., the mass parameter $\mu^2$ has been rewritten as $a\tau=a(1-T/T_c)$, with $a<0$. This means that the coefficient of the quadratic term in the free energy is a negative term that changes sign at the critical temperature $T>T_c$, where the material goes to the normal phase. In 3D materials, it was obtained by Gor'kov~\cite{Gorkov1958} $a=-N(0)$, where $N(0)=mk_F/2\pi^2\hbar^2$ is the density of states (DOS) at the Fermi level, $m$ is the electron mass and $\hbar k_F$ is the Fermi momentum. The parameter $\lambda$ has been rewritten as the GL coefficient $b=7\zeta(3)N(0)/8\pi^2T_c^2$, which is positive and temperature-independent and ensures that the free energy has a finite global minimum. Also, $\mathcal{K}=\hbar^2v_F^2b/6$ is the stiffness of the order parameter, which is proportional to the Fermi velocity, $v_F$, squared.
The field $\phi$, which from now on is defined as the order parameter of the system, $\Psi$, might be interpreted as a many-particle macroscopic wave function, as formulated originally by Ginzburg~\cite{Ginzburg:1950sr}. For uniform bulk materials in the absence of magnetic fields, the minimization of the free energy leads to:
\begin{equation}
\phi_u^2=-\frac{\mu^2}{2\lambda}>0,\qquad \mbox{or},\qquad \Psi_u^2=\frac{|a|\tau}{b}.
\end{equation}
We can see that the first two terms in the free energy functional are both of order $\mathcal{O}(\tau^2)$.

Furthermore, in the more general case, with an applied magnetic field and a non-uniform order parameter, minimization of the GL free energy functional leads to 
\begin{equation}\label{eq.OP}
a\tau\Psi-b|\Psi|^2\Psi-\mathcal{K}\mathbf{D}^2\Psi=0,
\end{equation}
where we defined the gauge-invariant derivative operator: $\mathbf{D}=\vec{\nabla}-i\frac{\hbar c}{2e}\mathbf{A}$. Note that the equation above can be re-written in dimensionless units, with $\Psi_u$ being the unit for the order parameter and
\begin{equation}
\xi = \sqrt{\frac{1}{|\, \mu^2\, |}},\qquad \mbox{or}, \qquad\xi=\sqrt{\frac{\mathcal{K}}{|a|\tau}},
\end{equation}
the coherence length, which is taken as the unit of length. It is divergent at $T_c$, with $\xi\propto\tau^{-1/2}$. 
Also, note that to maintain the same scaling in $\tau$ for all the terms in the free energy, one needs that the term $k\propto\tau^{1/2}$ and therefore the last term in Eq.~(\ref{Eq.GLfree}) is proportional to $\tau^2$ as all the others in the same equation. By doing so, one guarantees a consistent perturbation theory. As $k$ has dimensions of inverse length $(cm^{-1})$, then $k\to0$ as $T\to T_c$ which corresponds to a characteristic length that should to be divergent at $T\to T_c$. Note that in the case of the Fulde-Ferrel-Larkin-Ovchinikov (FFLO) states~\cite{Burkhardt1994}, the temperature dependence of the coherence length is not so important once it is supposed to exist at low temperatures. Notwithstanding, FFLO states are out of the scope of this article, once they violate time symmetry~\cite{Berg2009}. Furthermore, we do not consider temperature dependence of this term and rather express all quantities in temperature reduced dimensionless units.

Next, the minimization with respect to the vector potential leads to the current equation:
\begin{align}\label{eq.Curr}
\frac{1}{4\pi}\vec{\nabla}\times\vec{\nabla}\times\mathbf{A}&=i\mathcal{K}\frac{2e}{\hbar c}\left(\Psi^*\mathbf{D}\Psi-\Psi\mathbf{D}^*\Psi^*\right)\nonumber\\
&\qquad+\frac{1}{4\pi}\left[2k(\mathbf{x})\vec{\nabla}\times\mathbf{A}-\mathbf{A}\times\vec{\nabla}k(\mathbf{x})\right].
\end{align}
Or, in the case where $k$ is constant, we have simply:
\begin{equation}\label{eq.currconsK}
\frac{1}{4\pi}\vec{\nabla}\times\vec{\nabla}\times\mathbf{A}=i\mathcal{K}\frac{2e}{\hbar c}\left(\Psi^*\mathbf{D}\Psi-\Psi\mathbf{D}^*\Psi^*\right)+\frac{1}{2\pi}k\vec{\nabla}\times\mathbf{A}.
\end{equation}

In the case of a weak magnetic field applied to the superconductor, it is expected the Meissner effect, where the magnetic field is screened from the material. The modulus of the order parameter is not significantly reduced and we can consider its value approximately equal to the uniform solution, $\Psi(\mathbf{x})\approx\Psi_u$. In this case, one can neglect the terms involving the gradient of the order parameter, $\vec{\nabla}\Psi\approx0$, from Eq's.~(\ref{eq.Curr}) and~(\ref{eq.currconsK}) and obtain:
\begin{equation}
\frac{1}{4\pi}\vec{\nabla}\times\vec{\nabla}\times\mathbf{A}=2\mathcal{K}\left(\frac{2e}{\hbar c}\right)^2|\Psi_u|^2\mathbf{A}+\frac{1}{4\pi}\left[2k(\mathbf{x})\vec{\nabla}\times\mathbf{A}-\mathbf{A}\times\vec{\nabla}k(\mathbf{x})\right].
\end{equation}
and for the case of constant $k$:
\begin{equation}
\frac{1}{4\pi}\vec{\nabla}\times\vec{\nabla}\times\mathbf{A}=2\mathcal{K}\left(\frac{2e}{\hbar c}\right)^2|\Psi_u|^2\mathbf{A}+\frac{1}{2\pi}k\vec{\nabla}\times\mathbf{A}.
\end{equation}
In the standard GL theory, with $k=0$, the London penetration length, $\lambda_L$, i.e. the characteristic length of penetration for the magnetic field is extracted from the current equation by taking the rotor:
\begin{equation}
\nabla^2\mathbf{B}=\frac{1}{\lambda_L^2}\mathbf{B},\quad \mbox{with}\quad \lambda_L=\sqrt{\frac{b}{8\pi|a|\tau\mathcal{K}\left(\frac{2e}{\hbar c}\right)^2}}.
\end{equation}

In the case with $k\neq0$, it is possible to follow the same steps and get the modified London equation for the non-uniform case, $k(\mathbf{x})$:
\begin{align}
\nabla^2\mathbf{B}&=\frac{1}{\lambda_L^2}\mathbf{B}-2k\nab\times\mathbf{B}-2(\nab k)\times\mathbf{B}\nonumber\\
&+(\nabla^2k)\mathbf{A}-(\nab\cdot\mathbf{A})\nab k+(\nab k\cdot\nab)\mathbf{A}-(\mathbf{A}\cdot\nab)\nab k,\label{eq.nonconstK}
\end{align}
and also for the case with constant $k(\mathbf{x})=k$:
\begin{equation}
\nabla^2\mathbf{B}=\frac{1}{\lambda_L^2}\mathbf{B}+2k\vec{\nabla}\times\mathbf{B}.
\end{equation}
Finally, when we express distances in units of the London penetration length, $\mathbf{x}\to\lambda_L\tilde{\mathbf{x}}$ and $\vec{\nabla}\to\lambda_L^{-1}\tilde{\nabla}$, the equation above becomes:
\begin{equation}\label{eq.main}
\tilde{\nabla}^2\mathbf{B}=\mathbf{B}+2(k\lambda_L)\tilde{\nabla}\times\mathbf{B}.
\end{equation}
In this fashion, the theory is reduced to a single free parameter, $k\lambda_L$.

Systems with stratification of the order parameter could also be described in the lines shown above. What is typically done, is the introduction of a phase factor of the type $\Psi(\mathbf{x})\sim\Psi_ue^{i\mathbf{q}\cdot\mathbf{x}}$. It is easy to prove that one still gets the same current equation by taking the rotor on Eq.~(\ref{eq.Curr}), once the rotor operator introduces mixed derivatives which cancel the contributions from this phase factor $e^{i\mathbf{q}\cdot\mathbf{x}}$ and we still obtain equations similar to Eq's.~(\ref{eq.nonconstK}) and~(\ref{eq.main}).

The critical fields defined according to the standard criteria remain unchanged, namely, the thermodynamic critical field:
\begin{equation}\label{eq.Hc}
H_c=\sqrt{\frac{4\pi a^2\tau^2}{b^2}},
\end{equation}
and the critical field of the Abrikosov's vortices lattice:
\begin{equation}\label{eq.Hc2}
H_{c2}=\frac{\Phi_0}{2\pi\xi^2},
\end{equation}
where $\Phi_0=hc/2e$ is the quantum of flux~\cite{Abrikosov1957a}. Notwithstanding, it will be shown below that anomalous magnetic phases appear, leading to the need for a detailed study of the behavior of this system for different values of the parameter $k\lambda_L$.
Expressions~(\ref{eq.Hc}) and~(\ref{eq.Hc2}) give the temperature scalings of the critical fields near $T_c$, where both are $\propto\tau$, although we will not explore temperature aspects in this article and all the results below are valid for an arbitrary temperature.


We chose simple geometries to perform numerical simulations of the modified GL equations for a long cylindrical superconductor (with circular and square shapes) under the influence of a parallel magnetic field. We neglected variations along the axis of the field, say $z$-axis. In this case, even though all quantities depend on $x$ and $y$, not on $z$, we can have relevant values for the magnetic field and current in the $x$ and $y$ directions, which is not the main trend in usual superconductors. We employed a standard finite differences scheme with a square mesh. The details about the numerical methods are shown in the Supplemental Material file.

\begin{figure}
\centering
\includegraphics[width=\linewidth]{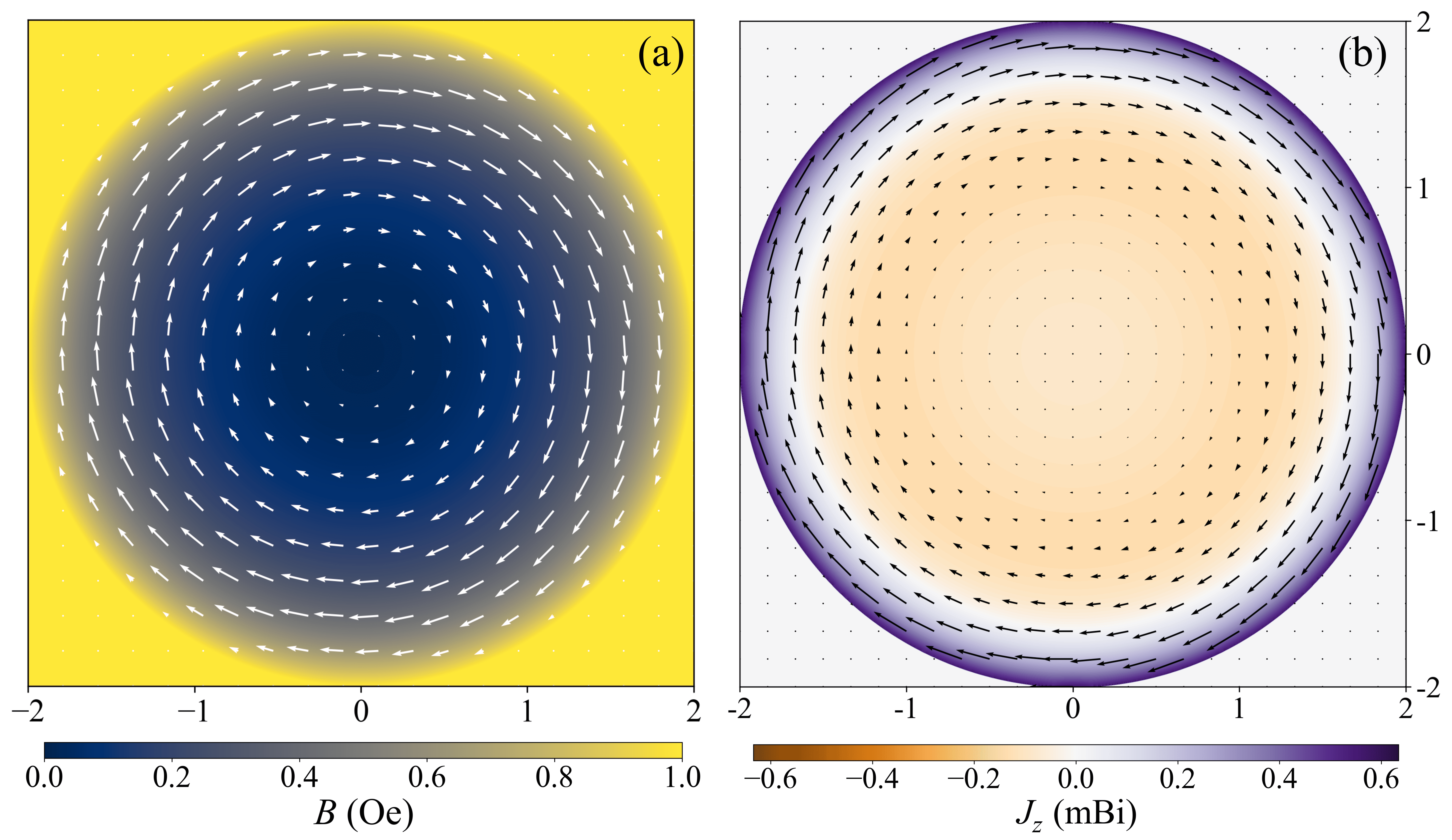}
\caption{\label{Fig1}
(a) Magnetic field intensity, $B$, for $k\lambda_L \approx 0.25$. The arrows indicate the magnetic field component in the $xy$-direction ($\mathbf{B}_{\parallel}$). The external field is applied solely in the perpendicular direction, with $B_z = 1$ Oe. 
(b) Corresponding current density distribution perpendicular to the circular cross-section of the circular sample. The London penetration depth is $\lambda_L = 400$ nm, and the cross-section of the long cylinder is circular with a radius of $2$ $\mu$m.}
\end{figure}

We found different phases of the magnetic response of the system depending on the dimensionless parameter $k\lambda_L$. In the case where this parameter is relatively small ($k\lambda_L\lesssim0.39$), as shown in Fig.~(\ref{Fig1}), solutions where the magnetic field points essentially in the $z$-direction, with extremely small components in the $xy$ plane. Also, the solution shows high surface currents which shield the interior of the sample from the external field, similarly as in the Meissner effect. Note that the longitudinal component of the current changes sign towards the center of the sample, still with extremely small value, around $\pm0.6\times10^{-3}Bi$.
We also found this type of solutions in square samples, as will be shown further. Moreover, the presence or not of corners at the edge of the sample does not cause any qualitative difference in behavior for small $k\lambda_L$. We explored a wide range of values for $k\lambda_L$, ranging from $10^{-6}$ up to $10^2$.

\begin{figure*}[p]
\centering
\includegraphics[width=0.99\linewidth]{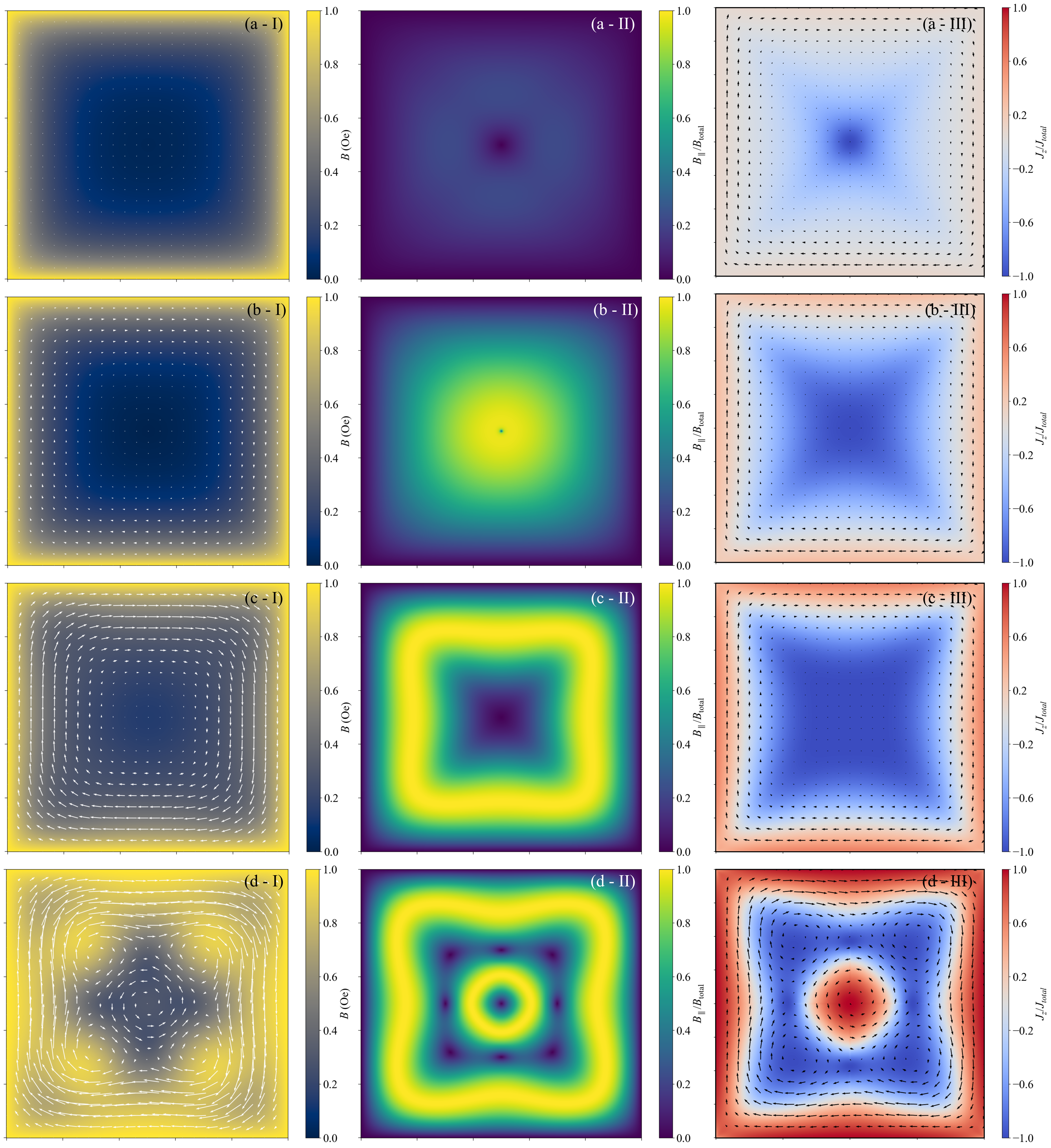}
\caption{\label{fig.matrix}The first column (I) presents a color map of the magnetic field intensity, $B$, with the in-plane components, \( B_{\parallel} \), represented by white arrows (in arbitrary units). The second column (II) shows the ratio between the magnetic field intensity in the cross-section and its total intensity. The plots in the third column (III) display the ratio of the electric current density in the \( z \)-direction, \( J_z \), to the total current density, \( J \). The current vectors in the $xy$-plane, $\mathbf{J}_\parallel = (J_x, J_y)$, normalized by the maximum absolute value of \( J_\parallel \) for \( k\lambda_L = 1 \). The figure represents a wire with a square cross-section of side length \( 10\lambda_L \), not a cylinder. The rows correspond to different values of \( k\lambda_L \) (0.1, 0.3162, 0.7943, and 1.0000).}
\end{figure*}

Above the value $k\lambda_L> 0.39$, the transversal component of the magnetic field becomes the total magnetic field inside the sample at a loop around the center.
Furthermore, we found the breakdown of the usual Meissner effect in the region with $k\lambda_L\approx 0.8$. As show in Fig.~(\ref{fig.matrix}),
one finds anomalous configurations above the threshold $k\lambda_L\gtrsim0.39$. For example, some strange vortex phases with large components of magnetic field perpendicular to the symmetry axis, i.e. the $xy$ direction, and also a different phase with an anti-vortex in the center. In Fig.~(\ref{fig.matrix}), in the rows (a) and (b) with $k\lambda_L=(0.1,0.3162)$, we found solutions similar to the Meissner effect. In row (c), with $k\lambda_L=0.7943$, there is a significant intensity of transversal components of magnetic field and currents. We see that transversal component of the magnetic field becomes the dominant component in a square-like stripe between the edges of the sample and the center. 
As the magnetic field in the $z$-direction penetrates the superconductor, it couples into the orthogonal components, leading to an increase in the $xy$ components. These components reach a peak before decreasing as the total magnetic field is progressively expelled from the material.
In row (d), an anti-vortex (a strong flux region pointing in the direction opposite to the external field with circular currents around it) appear in the center, while other four vortices stay in the diagonals of the square. The current configurations for $k\lambda_L\gtrsim0.8$ are also different from the screening Meissner currents because, instead of a circulating current around the vortices and surfaces, there is a significant component in the direction of the magnetic field. Besides that, the longitudinal component of the current changes sign, being positive near the edges and becoming negative near the center, as can be see in the third column of Fig.~(\ref{fig.matrix}). The larger the value of $k\lambda_L$ in the Eq.~(\ref{eq.main}), the stronger the coupling between the applied external field and the orthogonal components of the induced field inside the sample. In other words, the last term in Eq.~(\ref{eq.main}), determines how much of the total field is transferred to the orthogonal components. As seen in the last subplot, Fig.~(\ref{fig.matrix}), (d-III), The transversal components of the current circulate the zones of high field, namely four in the diagonals and one in the center. Again, four of these vortices have currents with clockwise rotation and one in the center with counter-clockwise rotation.



\begin{figure*}[p]
\centering
\includegraphics[width=\linewidth]{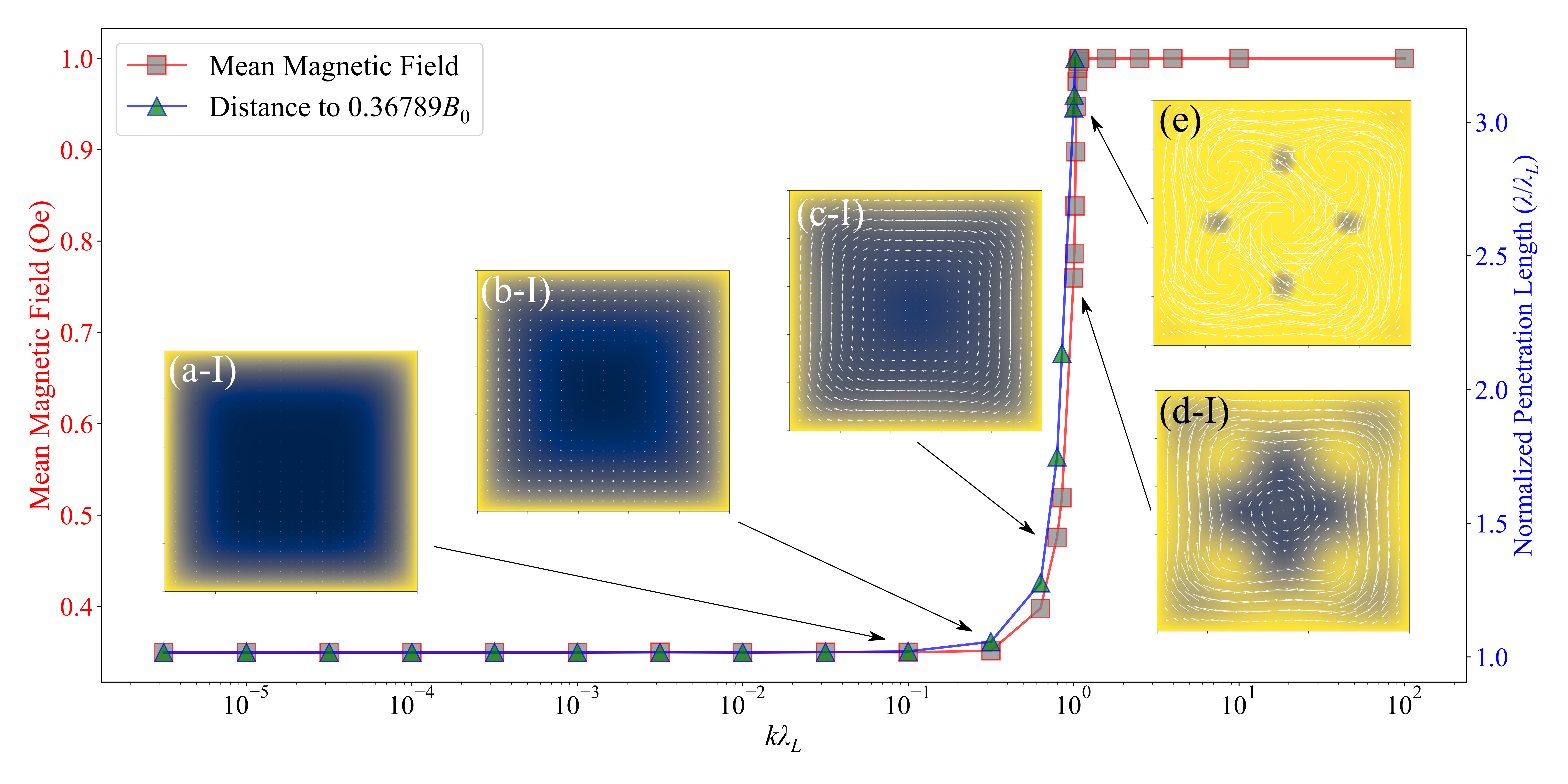}
    \caption{Dependence of the mean magnetic field (red curve, left scale) and the effective penetration length normalized by the London penetration length (blue curve, right scale) on $k\lambda_L$. The red curve represents the average magnetic field within the material as a function of $k\lambda_L$, while the blue curve shows the ratio of the effective penetration depth to the London penetration depth.}
    \label{Fig.PD}
\end{figure*}

We found a distinction of the magnetic signature of the QED inspired superconductor above a clear range of values of the CFJ-term. In order to quantify the difference of character between the Meissner limit for $k\lambda_L\to0$ and the new phases, we decided to calculate the mean value of the modulus of the magnetic field and the effective penetration length, defined as
\begin{equation}
\lambda_\text{eff} = \frac{1}{B_0}\int B\left(x\right) \, dx,
\end{equation}
where the integral is evaluated from the middle of the edge of the square until the center. As can be seen in Fig.~(\ref{Fig.PD}), both values increase sharply above $k\lambda_L\gtrsim10^{-1}$ and reach a limiting value for $k\lambda_L\gtrsim1$.
These exotic configurations, if reproduced experimentally, could be the hallmark for exotic states of unconventional Cooper-pairing mechanisms. For example, such models could play a major role in the scope of stratified superconductors such as UTe$_2$~\cite{Aoki2022}. It is a promising candidate for spin-triplet superconductivity, where the pairing mechanism is mediated by ferromagnetic fluctuations rather than the conventional electron-phonon interaction~\cite{Ran2019,Jiao2020,Knafo2021}. The potential violation of CPT-symmetry in such materials~\cite{Fu2015,Kozii2015,Schumann2020,Kanasugi2022} could lead to unusual phenomena, such as the emergence of Majorana fermions~\cite{Gardner2016,Kayser1983,Kayser1984,Feinberg1959}. In this case, the modulation of the order parameter happens in a particular direction of the space, generating a privileged frame of reference and thus breaking Lorentz invariance locally. Moreover, this non-standard formulation of the Ginzburg-Landau theory in systems with break-down of Lorentz invariance allows the possibility of having other forms of Bogomolny points within the GL formalism~\cite{Vagov2016,Saraiva2019}. In this case, the point of infinite degeneracy of the theory is shifted or even expanded into broader regions of the space of parameters. This is a point that deserves further investigation.

In summa, we investigated the different magnetic phases derived from a modified GL theory. The modification in the GL functional was the inclusion of a term inspired by the CFJ term, i.e. a term that violates Lorentz symmetry. In the static case, the equations that minimize the free energy produced equations similar to the GL equations with one extra term. We considered cases where the magnetic penetration length is much larger than the coherence length, namely the London limit, where one can neglect variations of the order parameter over the sample. The equation for the order parameter becomes trivial and one only has to solve the equation for the magnetic field. This equation became a kind of a modified London equation with an extra term. Our numerical solutions demonstrated that the magnetic response of superconductors becomes different from the usual Meissner effect with higher values of the scalar term arising from the CFJ term. There was a clear division in the phase-diagram where this term becomes large enough with respect to the London penetration length to produce a significant coupling of the longitudinal and transversal components of the magnetic field generating anomalous field configurations with rich phenomenology, such as anti-vortices and strong transversal magnetic fields.
This article opens a new theoretical framework of modeling superconductors with symmetry considerations.

\acknowledgments
M. C. Ara\'{u}jo would like to thank FUNCAP, Funda\c{c}\~{a}o Cearense de Apoio ao Desenvolvimento Cient\'{i}fico e Tecnol\'{o}gico (Process No. DC3-0235-00076.01.00/24), and CNPq, Conselho Nacional de Desenvolvimento Cient\'{i}fico  e Tecnol\'{o}gico - Brasil (Process No. 304145/2025-4) for financial support. TTS thanks the hospitality from the Federal University of Cear\'a, especially to Claudener Souza Teixeira and the financial support from Funda\c{c}\~ao Cearense de Apoio ao Desenvolvimento Cient\'{i}fico  e Tecnol\'{o}gico (FUNCAP) under grant number PVS-0215-00079.02.00/23. JF would like to thank Alexandra Elbakyan and Sci-Hub, for removing all barriers in the way of science and the Funda\c{c}\~{a}o Cearense de Apoio ao Desenvolvimento Cient\'{i}fico e Tecnol\'{o}gico (FUNCAP) under the grant PRONEM PNE0112- 00085.01.00/16 and the Conselho Nacional de Desenvolvimento Científico e Tecnol\'{o}gico (CNPq) under the grant 304485/2023-3. A.F.M would like to thank Alexandra Elbakyan and the Conselho Nacional de Desenvolvimento Científico e Tecnol\'{o}gico (CNPq) under the grant 104550/2025-2.


\bibliography{biblio}

\end{document}